\pdfoutput=1
\documentclass[twocolumn,amsmath,amssymb,prl,showpacs,superscriptaddress,preprintnumbers]{revtex4}

\usepackage{graphicx}
\usepackage{latexsym}
\usepackage{bm}
\usepackage{color}

\newcommand{\inputFMDG}[1]{\includegraphics{#1}}

\def\be{\begin{equation}}
\def\ee{\end{equation}}
\def\gs{\mathrel{
   \rlap{\raise 0.511ex \hbox{$>$}}{\lower 0.511ex \hbox{$\sim$}}}}
\def\ls{\mathrel{
   \rlap{\raise 0.511ex \hbox{$<$}}{\lower 0.511ex \hbox{$\sim$}}}}

\newcommand{\ba}{\begin{array}{c}}
\newcommand{\baz}{\begin{array}{cc}}
\newcommand{\barrr}{\begin{array}{rrr}}
\newcommand{\bad}{\begin{array}{ccc}}
\newcommand{\bav}{\begin{array}{cccc}}
\newcommand{\baf}{\begin{array}{ccccc}}
\newcommand{\bea}{\begin{equation} \begin{array}{c}}
\newcommand{\eea}{ \end{array} \end{equation}}
\newcommand{\beal}{\begin{equation} \begin{array}{l}}
\newcommand{\eeal}{ \end{array} \end{equation}}
\newcommand{\ea}{\end{array}}
\newcommand{\D}{\displaystyle}
\newcommand{\dms}{\mbox{$\Delta m^2_{\odot}$}}
\newcommand{\dma}{\mbox{$\Delta m^2_{\rm A}$}}
\newcommand{\meff}{\mbox{$\langle m \rangle$}}

\newcommand{\GeV}{\ensuremath{\,\mathrm{GeV}}}

\newcommand{\dmsq}{\Delta m^2}

\def\U2{\underline{U\hspace{-.9mm}}\hspace{.9mm}}

\newcommand{\dd}{\ensuremath{\mathrm{d}}}

\newcommand{\Figref}[1]{Fig.~\ref{#1}}

\begin{document}

\title{Lower Bounds on the Smallest Lepton Mixing Angle}

\preprint{IPPP/10/83; DCPT/10/166}

\author{Shamayita Ray}
\email[Electronic mail: ]{sr643@cornell.edu}
\affiliation{Laboratory for Elementary-Particle Physics (LEPP), Cornell University, Ithaca, NY 14853 USA}

\author{Werner Rodejohann}
\email[Electronic mail: ]{werner.rodejohann@mpi-hd.mpg.de}
\affiliation{Max--Planck--Institut f\"ur Kernphysik, 
Postfach 103980, D--69029 Heidelberg, Germany}

\author{Michael A.~Schmidt}
\email[Electronic mail: ]{m.a.schmidt@durham.ac.uk}
\affiliation{Institute for Particle Physics Phenomenology (IPPP),
Durham University, Durham DH1 3LE, UK}


\begin{abstract} 
\noindent
We give minimal values for the smallest lepton mixing parameter 
$|U_{e3}|$, applying 2-loop renormalization group equations in an 
effective theory approach. This is relevant in scenarios 
that predict an inverted neutrino mass spectrum with the smallest mass
and $|U_{e3}|$ being zero at tree level, a situation
known to be preserved at 1-loop order. 
At 2-loop, $|U_{e3}|$ is generated at a level of  
$10^{-12}$--$10^{-14}$. Such small values are of interest in 
supernova physics. Corresponding limits for the normal mass ordering 
are several orders of magnitude larger. 
Our results show that $|U_{e3}|$ can in general expected be to
be non-zero.

\end{abstract}

\pacs{14.60.Pq; 11.10.Gh}

\maketitle

The subject of lepton mixing and neutrino mass has entered the
precision era. 
It is confirmed that there are three flavors of 
neutrinos which mix among themselves because the flavor states are not
identical to the states with distinct masses $m_1$, $m_2$ and $m_3$.
The parameters describing the mixing are the three 
angles $\theta_{12}$, $\theta_{23}$, $\theta_{13}$ and a 
phase $\delta$. Two more phases $\varphi_{1,2}$ are needed 
if neutrinos are Majorana particles, which is the case in basically
all extensions of the Standard Model, and also from an effective 
field theory point of view. 
 The two mixing angles $\theta_{12}$ and $\theta_{23}$ as well
  as the mass squared difference 
$\dmsq_{\odot} \equiv m_2^2 - m_1^2$ and the magnitude of 
$\dmsq_{\rm A} \equiv m_3^2 - m_2^2$ have been determined with high 
accuracy, and further improvement is foreseen with a variety 
of experimental approaches \cite{rev_ex}.
However, of particular interest is the parameter 
$|U_{e3}| = \sin \theta_{13}$, which describes the 
electron neutrino content in the heaviest (lightest) 
neutrino mass-eigenstate, if the normal (inverted) 
mass ordering is realized, {\it i.e.}~if $\dmsq_{\rm A} > 0$ 
($\dmsq_{\rm A} < 0$). 
The main feature of $|U_{e3}|$ is its smallness compared 
to the other mixing angles which 
makes it difficult to be measured experimentally \cite{rev-th13}. 
Three independent groups performing global fits of the 
world's neutrino data find $3\sigma$ limits of 
$|U_{e3}|^2 \le 0.046$ \cite{fogli}, $\le 0.053$ \cite{VS}, 
and $\le 0.043~(0.047)$ \cite{GM}, 
where the two values for Ref.~\cite{GM} stem from two slightly
different analyses of solar data. 
$|U_{e3}|$ is of fundamental importance as 
it is crucial to rule out a large number of models
constructed to explain the peculiar structure of lepton
mixing \cite{AC}, CP violation in oscillations 
depends crucially on $|U_{e3}|$, and it also plays an important role 
in neutrino-less double beta decay \cite{LMR}.  
Although some rather weak hints on a non-zero value exist  
\cite{fogli} at $\sim$1.5$\sigma$, all existing data are well 
compatible with $|U_{e3}| = 0$.   

The smallness of $|U_{e3}|$ is most naturally explained by
  some flavor symmetry~\cite{Ue30_mod,scal0,scal} predicting
  $|U_{e3}|$ to vanish at some high scale $\Lambda$. In general, the symmetry 
is broken and potentially generates small but non-zero values of $|U_{e3}|$ 
at the low scale $\lambda$ relevant for the oscillation 
experiments. Radiative corrections are one inevitable source
  of this breaking. In this letter we discuss the
  minimally induced values of $|U_{e3}|$ through renormalization group (RG) 
evolution, in order to investigate whether non-zero values of
$|U_{e3}|$ can be expected in general. 
 Given the huge amount of experimental activity and the theoretical
importance of this parameter, it is obviously of interest to 
investigate whether non-zero values can be expected on general
grounds.  
 We therefore especially focus on the case of an inverted mass 
 ordering with $m_3=|U_{e3}|=0$, which is a fixed point at 1-loop. It
is shown that at 2-loop order this is no longer a fixed point, and we
argue that it leads to the lowest possible
  value of $|U_{e3}|$, if extreme fine-tuning of the parameters is
neglected. The order of magnitude of $|U_{e3}|$ at low scale lies 
between $10^{-12}$ and $10^{-14}$. This is, of course, beyond the
reach of the planned neutrino oscillation facilities, but of interest in 
supernova physics. It is furthermore easy to rule out this
possibility. 
We also give the corresponding bounds on $|U_{e3}|$ for the normal
mass-ordering, which happens to be several orders of magnitude larger.

The 1-loop RG evolution of $\theta_{13} = \sin^{-1} |U_{e3}|$, up to
first order, is given by \cite{L,RG_rev}
\begin{eqnarray}
&& \dot \theta_{13} =  \frac{C \, y_\tau^2}{32 \pi^2} \, \sin 2 \theta_{12}
\, \sin 2 \theta_{23} \frac{m_3}{\Delta m^2_{\rm A} \, (1 + R)} \times  \label{eq:dot13}\\
&& \left[ m_1 \cos (\varphi_1 - \delta) - (1 + R) m_2 \cos (\varphi_2 -
\delta) - R  m_3 \cos \delta \right]  \, , \nonumber 
\end{eqnarray}
where the dot denotes d/d$t = \mu\,$d/d$\mu$ 
with $t \equiv \ln{(\mu/\Lambda)}$, $\mu$ being the renormalization scale. 
One has $C = -\frac 32$ $(C=1)$ in the Standard Model (SM) 
(Minimal Supersymmetric SM (MSSM)),
$R = \dms/\dma$,  
and the tau Yukawa coupling is $y_\tau =
\sqrt{2} \, (m_\tau/{\rm GeV}) /246 \simeq 0.010$ in the SM and 
$0.010 \, (1 + \tan^2 \beta)$ in the MSSM. 
The electron and muon Yukawa couplings $y_{e, \mu}$ have been
neglected here. 
Eq.~(\ref{eq:dot13}) shows that the evolution of $|U_{e3}|$ 
is proportional to $m_3$. Moreover at 1-loop, one has 
$\dot m_i \propto m_i \, (i=1,2,3)$ 
and thus $|U_{e3}|(\mu) = m_3(\mu) = 0$
is a fixed point of RG evolution: these quantities remain zero
throughout the RG evolution. 
 The stability of this scenario under radiative corrections remains 
true even if $y_{e, \mu}$ are taken into account \cite{scal0}.

To understand the existence of the fixed point in a more general form, 
we consider the neutrino mass matrix $m_\nu$ for $m_3 = |U_{e3}| = 0$, 
which obeys a structure denoted as ``scaling'' \cite{scal}: 
\be
m_\nu^{\rm {scaling}} = \left( \bad 
a & b & b/c \\
\cdot & d & d/c \\
\cdot & \cdot & d/c^2 
\ea \right) \; , 
\ee
where $\tan^2 \theta_{23} = |1/c^2|$ and $a$, $b$, $d$ 
are determined by $m_{1,2}$, 
$\theta_{12}$, and the 
Majorana phase difference $\Delta \varphi = \varphi_2 - \varphi_1$. 
At 1-loop, the RG equation of $m_\nu$ is given as \cite{L,RG_rev}
\be \label{eq:mnuRG1}
\dot m_\nu = \frac{1}{16 \pi^2} \left[ 
\alpha_\nu \, m_\nu + C \left( Y^T \, m_\nu + m_\nu \, Y \right)
\right] , 
\ee
where $Y = Y_\ell \, Y_\ell^\dagger$ with 
$Y_\ell = {\rm diag}(y_e, y_\mu, y_\tau)$ being the charged lepton Yukawa matrix.
The parameter $\alpha_\nu$ is a function of gauge and 
Yukawa couplings (as well as the Higgs self-couplings in the SM), 
which does not lead to modifications of the mixing matrix elements. 
From Eq.~(\ref{eq:mnuRG1}) RG evolution of an element $(m_\nu)_{\alpha \beta}$ 
becomes \cite{EL}
\be
\dot{(m_\nu)}_{\alpha \beta} = \frac{1}{16 \pi^2} \left[ \alpha_\nu + 
C \left( y_\alpha^2 + y_\beta^2 \right) \right] (m_\nu)_{\alpha \beta}
\, , 
\ee
which at low scale leads to the RG-modified mass matrix 
\begin{eqnarray}
m_\nu = I_{\alpha_\nu}\left( 
\begin{array}{ccc}
(m^0_\nu)_{ee} \, I_e^2  & (m^0_\nu)_{e\mu} \, 
I_e \, I_\mu  & (m^0_\nu)_{e\tau} \, I_e \, I_\tau \\ 
\cdot & (m^0_\nu)_{\mu\mu} \, I_\mu^2  & (m^0_\nu)_{\mu\tau} \, I_\mu \, I_\tau \\ 
\cdot & \cdot & (m^0_\nu)_{\tau\tau} \, I_\tau^2 
\end{array} 
\right) , \quad \label{eq:mnu1}
\end{eqnarray}
where $m^0_\nu$ denotes the mass-matrix at the high-scale and 
\bea
I_\alpha  = \exp \left( \frac{C}{16 \pi^2} 
\int_\Lambda^\lambda y_\alpha^2 \, {\rm d}t \right ) 
\label{Ialpha},
\eea
with $\alpha, \beta \in \{ e, \mu, \tau \}$. 
Eq.~(\ref{eq:mnu1}) shows that the scaling property of
$m_\nu^{\rm scaling}$ is preserved by the RG evolution since the
second and third column remain proportional to each other. 
The atmospheric neutrino mixing changes to $\tan^2 \theta_{23} 
= I_\tau^2/(I_\mu^2 \, |c^2|)$. Thus 
$m_3 = |U_{e3}| = 0$ remains valid at all energies during 1-loop
running. Therefore, any non-zero $|U_{e3}|$ generated at 2-loop can be
considered to be the smallest guaranteed value for $|U_{e3}|$, which 
obviously is of interest in its own right, and 
should  in the end represent the final precision goal for any
search. If this 2-loop value cannot vanish, it follows that in general
 $|U_{e3}|$ can be expected to be non-zero.  

The largest contribution in Eq.~(\ref{Ialpha}) will come from 
$|C| \, y_\tau^2/(16 \pi^2) \simeq 9.5 \times 10^{-7}$ 
for the SM and $\simeq 6.3 \times 10^{-7} \, (1+\tan^2\beta)$ for the
MSSM. $I_\alpha$ in Eq.~(\ref{Ialpha}) can thus be approximated as 
$I_\alpha  \simeq 1 +  \frac{C}{16 \pi^2} \, y_\alpha^2 \, \ln \frac{\lambda}{\Lambda} 
\equiv 1 + \epsilon_\alpha $.\\

\begin{figure}
 \scalebox{0.8}{\inputFMDG{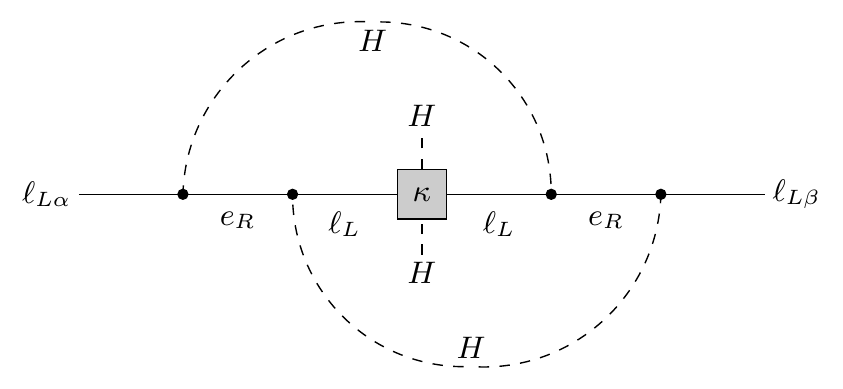}}
\caption{Rank-changing 2-loop diagram in the SM which breaks the 
1-loop RG invariance $m_3=|U_{e3}|=0$. $\ell_L$ ($e_R$) 
are left-(right-)handed leptons and $H$ is
the Higgs doublet.
}\label{fig:2loop}
\end{figure}

At 2-loop, the relevant term in the RG evolution equation of $m_\nu$ in 
case of the SM is \cite{smallest}
\begin{equation} \label{eq:mnuRG2}
\dot m_\nu = \frac{2}{(16 \pi^2)^2} \, 
Y^T \,  m_\nu \, Y \, ,
\end{equation}
which arises from the diagram shown in \Figref{fig:2loop}.
In the MSSM, there are no such terms 
(because of the non-renormalization theorem) 
unless supersymmetry (SUSY) is broken ~\cite{2loop}. 
Hence in the remaining part of this paper we will 
consider the SM, unless otherwise specified.  
From Eq.~(\ref{eq:mnuRG2}) the evolution of 
${(m_\nu)}_{\alpha \beta}$ can be written as
\be
\dot{(m_\nu)}_{\alpha \beta} = \frac{2}{(16 \pi^2)^2} 
\, y_\alpha^2 \, y_\beta^2 \, (m_\nu)_{\alpha \beta}  \, , 
\ee
and the mass-matrix $m_\nu$ gets modified to 
\be \label{eq:mnu2}
m_\nu \propto
\left( \bad 
(m^0_\nu)_{ee} \, I_e^2 \, I_{ee}  & (m^0_\nu)_{e\mu} \, I_e \, I_\mu \,
I_{e\mu} & (m^0_\nu)_{e\tau} \, I_e \, I_\tau \, I_{e \tau} \\ 
\cdot & (m^0_\nu)_{\mu\mu} \, I_\mu^2 \, I_{\mu\mu} 
& (m^0_\nu)_{\mu\tau} \, I_\mu \, I_\tau \, I_{\mu\tau} \\ 
\cdot & \cdot & (m^0_\nu)_{\tau\tau} \, I_\tau^2   \, I_{\tau\tau}
\ea \right) \, . 
\ee
The new parameters $I_{\alpha \beta}$ are given by 
\bea \label{I-2loop}
I_{\alpha\beta} = \exp \left\{ 
\frac{2}{(16 \pi^2)^2} \int y_\alpha^2 \, y_\beta^2\, {\rm d}t \right
\} 
\\ \simeq 
1 +  \frac{2}{(16 \pi^2)^2} \, y_\alpha^2 \, y_\beta^2 \, 
\ln \frac{\lambda}{\Lambda} 
\equiv 1 + \epsilon_{\alpha\beta} \, .
\eea
As can be seen, 2-loop running increases the rank of $m_\nu$ from 2 to 3.
In Eq.~(\ref{I-2loop}) the largest contribution comes from  
$|\epsilon_{\tau\tau}| \simeq 2.0 \times 10^{-11}$, 
and thus numerically is of the same order as that of $\epsilon_e$ at 1-loop. 
Here we have taken $\lambda = 10^2$ GeV and 
$\Lambda = 10^{12}$ GeV, which will be used throughout this paper. 
Obviously, the scale of $\epsilon_{\tau\tau}$ sets the value
for $|U_{e3}|$ generated at 2-loop order. 

To explicitly calculate the implied smallest value of $|U_{e3}|$ after 
2-loop running, we consider the scaling scenario $m_3=|U_{e3}|=0$ 
at the high scale $\Lambda$. It suffices to take only 
$\epsilon_{\tau\tau}$ into account and neglect all other 
$\epsilon_{\alpha\beta}$ in Eq.~(\ref{I-2loop}). 
As a result of the change in rank of $m_\nu$, 
the 2-loop running, unlike the 1-loop case, 
gives rise to a non-zero smallest neutrino mass 
\bea \nonumber 
m_3 = \frac{1}{4} \, \sin^2 2 \theta_{23} 
\left( m_2 \, \cos^2 \theta_{12} \, e^{i \Delta \varphi} + 
\sin^2 \theta_{12} \, m_1 \right)  |\epsilon_{\tau\tau}| \\ \nonumber 
\simeq \frac{1}{4} \, m_2 \, \sin^2 2 \theta_{23}  
\left(  \cos^2 \theta_{12} \, e^{i \Delta \varphi} + 
(1 + \frac{R}{2} ) \, \sin^2 \theta_{12} \right)  |\epsilon_{\tau\tau}| \, . 
\eea
The order of 
$m_3$ is $\simeq m_2 \, |\epsilon_{\tau\tau}|/4 
\sim 2.5 \times 10^{-13}$ eV for $m_2 \simeq 0.05$ eV. 
It confirms the result from Ref.~\cite{smallest}.

\begin{figure}
\begin{center}
\includegraphics[width=\linewidth]{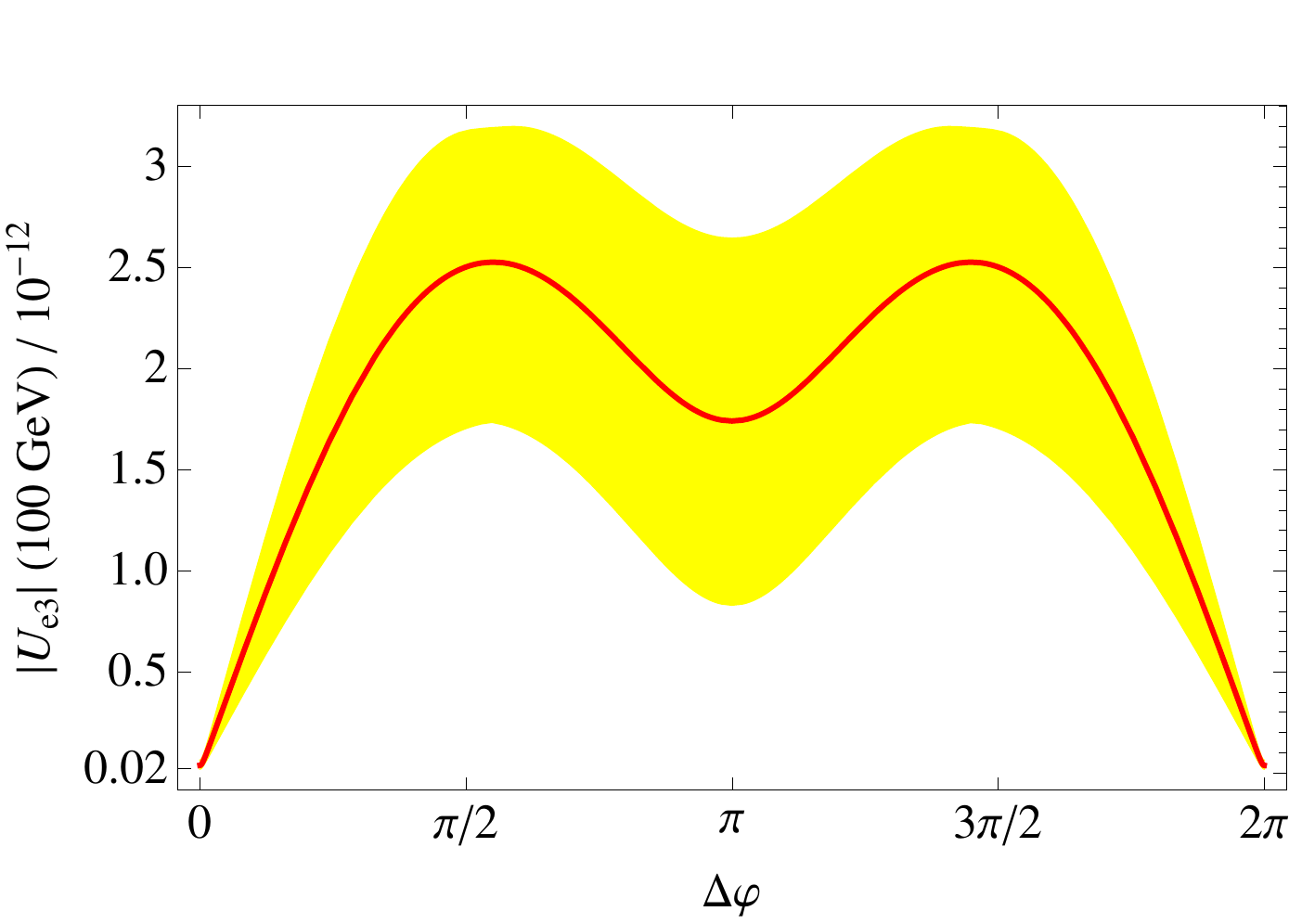}
\caption{Generated value of $|U_{e3}|$ at $\lambda=100\GeV$ assuming a 
vanishing value at $\Lambda=10^{12}\GeV$ in the 
leading-log approximation. The (solid) red line corresponds to fixing all 
mixing parameters to the best-fit 
values and the yellow region shows the $3\sigma$ region, which are 
taken from~\cite{VS}.}\label{fig:DepOnPhases}
\end{center}
\end{figure}

The 2-loop evolution equation for $|U_{e3}|$ in the scaling scenario 
can be obtained from Eq.~(\ref{eq:mnuRG2}), 
following the procedure of \cite{pol} (see also \cite{S}), to be
\bea \D \label{eq:main}
\frac{\dd}{\dd t}|U_{e3}|\simeq |\dot{U}_{e3}|\simeq \frac{y_\tau^4}{2 \, (16 \pi^2)^2} 
\frac{s_{23}^2}{m_1 \, m_2} \sin 2 \theta_{12} \, \sin 2 \theta_{23} \times \\
\D 
\bigl| m_1 - m_2 \, e^{i \Delta \varphi}\bigr| \, 
\bigl| m_1 \, s_{12}^2 + m_2 \, c_{12}^2 \, e^{i \Delta \varphi}\bigr| \; ,
\eea 
in the limit $y_{e,\mu}\ll y_\tau$. The
$\beta$-function of $|U_{e3}|$ does not depend on the Dirac CP phase
$\delta$ for vanishing $|U_{e3}|$ \cite{L}, which has been discussed
in detail in \cite{theta13Null}.
From Eq.~(\ref{eq:main}), the typical order of magnitude of
$|U_{e3}|$ at low scale $\lambda$ 
is $|U_{e3}| \simeq 
y_\tau^4 /(4 \, (16 \pi^2)^2) \, \ln\Lambda/\lambda \simeq 10^{-12}$. 
Fig.~\ref{fig:DepOnPhases} shows the dependence of
$|U_{e3}|$ on the Majorana phase difference $\Delta \varphi$. 
The maximal value of $|U_{e3}|$ is obtained for 
$\cos\Delta\varphi = (\tan^2 \theta_{12} - 1) 
(m_2^2 - m_1^2 \, \tan^2 \theta_{12})/(4 \, m_1 \, m_2 \, 
\tan^2 \theta_{12})$,  
and gives 
\be \label{eq:max}
|U_{e3}| \simeq 3.2 \times 10^{-12} \, , 
\ee 
while the minimum value, obtained for $\Delta\varphi=0$, is 
\be \label{eq:min}
|U_{e3}| \simeq 2.0 \times 10^{-14} \, . 
\ee 
It is suppressed with respect to Eq.~(\ref{eq:max}) 
by a factor of order $R \simeq 1/30$, which remains 
true even if $y_{e,\mu}$ are taken into account. 
It is clear that a 3-loop effect could not cancel this 
suppressed value, since it would be further suppressed 
by $1/(16 \pi^2) \ll R$.

One possibility for a smaller value of $|U_{e3}|$ would be if an 
initially non-zero value accidentally runs to (almost) zero in the
course of its evolution which, of course, will need an extreme amount
of fine-tuning.
Another possibility arises when the high energy completion of 
the effective theory is considered and the RG effects in the 
complete theory conspire with the effective running to give 
tiny values, {\it e.g.}~threshold effects in case of seesaw models. 
However, that will depend on the particular high 
energy extension considered, and will again involve extreme 
fine-tuning. For example, in seesaw models threshold effects 
generally lead to larger RG running than the running in the 
effective theory \cite{S}, and hence a fine-tuned cancellation 
will be required to produce the tiny $|U_{e3}|$ of relevance in 
this discussion. Hence it is natural to claim that the 
2-loop generated $|U_{e3}|$ as given in Eq.~(\ref{eq:main}) 
is the smallest value one should expect. Moreover, it shows that
non-zero $|U_{e3}|$ can be expected on general grounds.

As previously mentioned, $|U_{e3}|$ is not generated at 2-loop in
the MSSM, as long as SUSY is unbroken. However, analogous to
the discussion for $m_3$ in \cite{smallest}, $|U_{e3}|$ is
generated as SUSY is broken and receives 
contributions of two different origins. First, there
is a 2-loop contribution with sleptons and selectrons in the loop as 
shown in Fig.~1 of \cite{smallest}. Although it is not enhanced by 
a large logarithm, it depends on $(1 + \tan^2\beta)^2$ along with a 
complicated order one function of sparticle masses. Second, there 
can be corrections due to the 
the SUSY breaking operator $\tilde{L} H_u \tilde{L} H_u$~\cite{SUSY2}, 
which leads to $|U_{e3}|$ of similar order as the
2-loop contribution. Thus while a general prediction is not 
possible, it suffices to estimate the order of magnitude to be 
\be
|U_{e3}|^{\rm MSSM} \sim |U_{e3}|^{\rm SM} \, (1 + \tan^2\beta)^2 \, /
\ln \frac{\Lambda}{\lambda} \, , 
\ee
where $|U_{e3}|^{\rm SM}$ corresponds to the values in the SM 
discussed above. Thus even for moderate $\tan\beta$, 
$|U_{e3}|^{\rm MSSM}$ will be larger than the corresponding SM value.

Can such values be tested? With the present view, 
the ultimate experiment in order to study  
neutrino oscillations will be a neutrino factory. However, 
the facilities currently under study \cite{rev_ex} have a 
$3\sigma$ discovery potential on $|U_{e3}|$ of at most 
$1.5 \times 10^{-3}$, and hence the 2-loop generated 
tiny $|U_{e3}|$ is far beyond its scope. 

However, tiny non-zero values of $|U_{e3}|$, as discussed here, may have 
observable consequences in the neutrino spectra emitted by 
supernovae. Extremely high neutrino densities around the 
neutrinosphere give rise to so-called collective effects \cite{revs}, 
of which ``spectral swapping'' is the one where $\nu_e$ and 
$\nu_{\mu,\tau}$ swap their energy spectra at high (low) energies but 
keep their spectra at low (high) energies in case of inverted (normal) 
hierarchy. Here ``high'' and ``low'' energies are understood relative to 
a critical energy $E_C$, which decreases with $|U_{e3}|$ 
in the normal ordering, while is independent of 
$|U_{e3}|$ if the mass-ordering is inverted \cite{swap}.
The effect can be well understood with the analogy of 
an inverted pendulum \cite{pend}, which turns around by the 
slightest instability, generated by a non-zero $|U_{e3}|$ in 
this case. In case of inverted mass-ordering, 
the spectral swap occurs regardless of how small 
but non-zero $|U_{e3}|$ is, the impact of $|U_{e3}|$ being 
only to introduce a logarithmic dependence on the radius at which 
the conversion sets in. Ref.~\cite{swap} discusses in this context 
values of $|U_{e3}|$  
down to $10^{-70}$, 56 orders of magnitude smaller than the lower
limit obtained here.  

While the discussion here may be a bit too simplified
(the effects may depend on supernova details, they can be 
also induced by other effects like an asymmetry in the primary 
$\nu_\mu$--$\nu_\tau$ fluxes as well as radiatively induced
  matter effects for $\nu_{\mu,\tau}$~\cite{SN-trigger}, 
etc.), it catches the main point
to be made here: tiny values of $|U_{e3}|$, in particular the
ones generated by 2-loop running in the inverted hierarchy,
are expected to have observable consequences in
supernova physics. The lower bounds on $|U_{e3}|$ presented
here may be helpful in analyses of supernova neutrino analyses.

The scenario leading to the smallest possible value of $|U_{e3}|$ can
easily be tested or ruled out, not only by long-baseline oscillation
experiments pinning down the mass ordering, but also 
via neutrino mass related observables: for $m_3 = |U_{e3}| = 0$ 
the effective mass governing neutrino-less double beta decay is simply 
\be
\meff = \sqrt{|\dma|} \, \sqrt{1 - \sin^2 2\theta_{12} \, 
\sin^2 \Delta \varphi/2} \, . 
\ee
The Majorana phase difference $\Delta \varphi$ is the same as the 
one determining the magnitude of $|U_{e3}|$ and is, in principle, 
measurable by precise measurements of double beta decay \cite{CP}. 
In contrast to its value in the normal ordering, 
$\meff$ is bounded from below by $\sqrt{|\dma|} \, \cos 2 \theta_{12}$, 
which is testable in future experiments \cite{0vbb}. 
In what regards cosmology, the sum of neutrino
masses is about $2\sqrt{|\dma|} \simeq 0.1$ eV, which is twice as high
as its value in the normal ordering, and might be
testable in future measurements as well \cite{cosmo}.

In case of normal ordering the minimal $|U_{e3}|$ 
can trivially be obtained from the 1-loop running given 
in Eq.~(\ref{eq:dot13}). The minimal value is 
obtained with $\varphi_1 = \varphi_2 = \pi$ to be
\be 
|U_{e3}| = \frac{|C| \, y_\tau^2}{32 \pi^2}  \sin 2 \theta_{12} \, 
\sin 2 \theta_{23} \frac{m_3 \, (m_2 - m_1)}{(m_1 + m_3)\, (m_2 + m_3)} \,  
\ln \frac{\Lambda}{\lambda}\,,
\ee 
which  quantifies to $|U_{e3}| \sim 1.1 \times 10^{-6}$ 
in the SM and $\sim 1.6 \times 10^{-6} \, (1 + \tan^2 \beta)$ in 
the MSSM. Interestingly,  $|U_{e3}|$ is smallest for quasi-degenerate
neutrinos and not for $m_1 = 0$. 
Introducing the common mass scale $m_0$ of the quasi-degenerate
neutrinos, the lowest values are decreased by a factor of 
\[ \sim \frac{8 m_0^2}{\left( \sqrt{\dms} + \sqrt{\dma}\right) \sqrt{\dms}} 
\simeq 1600 \left(\frac{m_0}{0.3 \, \rm eV}\right)^2 \]
and hence reduce the scale of $|U_{e3}|$ by at most three 
orders of magnitude, and still stay three orders above the 
2-loop value in the inverted hierarchy.\\

In summary, we have analyzed the 2-loop running of $U_{e3}$ in the 
effective low energy theory and showed that it gives the minimal 
value of $|U_{e3}|$ in scenarios of inverted mass 
hierarchy with $U_{e3}=0$ at high scale. We focussed on this
scenario because at 1-loop there is no generation of a non-zero value,
and because it is of interest to investigate whether non-zero $|U_{e3}|$ can be
expected in general. 
The order of magnitude of $|U_{e3}|$ generated in this way 
is between $10^{-12}$ and $10^{-14}$ depending on the Majorana
phases. Though such small values are beyond the reach of currently
foreseen neutrino oscillation experiments, they are obviously 
of fundamental interest, and possess applications in supernova physics.\\


\noindent
{\sl Acknowledgments}:
The work of W.R.~was supported by the ERC under the Starting Grant 
MANITOP and by the DFG in the project RO 2516/4-1 as well as in the 
Transregio 27. S.R.~would like to thank Prof.~Yuval Grossman for 
support and hospitality. M.A.S.~expresses his gratitude to 
Manfred Lindner for hospitality at the MPIK.  


\end{document}